\documentclass{sig-alternate-10pt}

\usepackage{subfigure}
\usepackage{paralist}
\usepackage{balance}
\usepackage{hyperref}
\usepackage{breakurl}

\begin{document}

\title{Towards Defeating the {\ttlit Crossfire} Attack using SDN}
%
%
%
%
%

\numberofauthors{3} 
%
\author{
%
%
\alignauthor
Dimitrios Gkounis\\
       \affaddr{FORTH, Greece}\\
       \email{dgkounis@ics.forth.gr}
\alignauthor
Vasileios Kotronis\\
       \affaddr{ETH Zurich, Switzerland}\\
       \email{vkotroni@tik.ee.ethz.ch}
\alignauthor 
Xenofontas Dimitropoulos\\
       \affaddr{FORTH, Greece}\\
       \email{fontas@ics.forth.gr}
}

\maketitle
\begin{abstract} 

In this work, we propose online traffic engineering as a novel approach 
to detect and mitigate an emerging class of stealthy Denial of Service (DoS)
link-flooding attacks. Our approach exploits the Software Defined Networking
(SDN) paradigm, which renders the management of network traffic more flexible
through  centralised flow-level control and monitoring.  We implement a
full prototype of our solution on an emulated SDN environment using 
OpenFlow to interface with the network devices. We further discuss
useful insights gained from our preliminary experiments as well as a
number of open research questions which constitute work in progress.


\end{abstract}




\section{Introduction}


Some of the most powerful DoS attacks ever, such as the attack against Spamhaus which reached 400 Gbit/s~\cite{spamhaus}, 
have been observed during the last two years. Moreover, new types of DoS link-flooding attacks have been recently proposed; these attacks 
are extremely difficult to mitigate and could potentially take entire countries off the Internet~\cite{crossfire,Coremelt}. 
The \textit{Crossfire}~\cite{crossfire} attack in particular:
\begin{inparaenum}[\itshape i\upshape)]
\item uses legitimate rather than spoofed source IP addresses to send traffic; these addresses are much harder to filter,
\item sends legitimate packets to publicly accessible servers, and
\item transmits low-bandwidth flows from each bot individually; these ``under-the-radar'' flows then cumulatively flood certain
links in the network. 
\end{inparaenum}


In this work, we introduce a novel approach to detect \textit{Crossfire} bots under adverse DoS conditions. 
The key idea is to repeatedly change how traffic is routed and to monitor sources that persistently react
to re-routing for the purpose of attacking a specific target. If certain sources are recorded several times
in links that are DoS'ed they are considered suspicious and are filtered. Therefore, our approach effectively
increases the cost for executing \textit{Crossfire}.
Traditionally, traffic engineering is used solely for the purpose of load-balancing DoS attacks. We use online
traffic engineering both to detect stealthy bots and to balance the load. Our approach benefits from the
following SDN principles:
\begin{inparaenum}[\itshape i\upshape)]
\item centralised visibility and control for fast traffic engineering, and 
\item flow-level reactive traffic management.
\end{inparaenum}


\section{Methodology}

\textbf{Attacker Model:} The goal of the \textit{Crossfire}~\cite{crossfire} attack is to cut-off Internet connectivity towards a specific geographic 
area, the \textit{Target Area}. Around the target area, there are public servers, the \textit{Decoy Servers}, and network
links, the \textit{Target Links}, which lead to both the target area and the decoy servers. The attacker first constructs a
map of the links (the \textit{link-map}) in and around the target area. Then, he floods certain target links sending traffic only to the decoy
servers. This way the flood cannot be directly observed in the target area, but it still isolates this area from the rest of the network. 
For this purpose, the attacker finds the target links
that are used most along the bot-to-decoy server and bot-to-target area paths and he coordinates his bots to attack them. 
Finally, the attacker monitors the network routes and reacts to changes (which are possible defender's actions) by setting
up the attack again, i.e., updating the link-map and recalculating the target links.


\textbf{Defender Model:} The goal of the defender is to keep the network running 
without any severe congestion and to find and rate-limit malicious traffic sources.
Therefore, the defender: 
\begin{inparaenum}[\itshape i\upshape)]
\item monitors traffic load on his network and detects DoS'ed links that are severely congested,
\item balances traffic load by rerouting traffic destined to different destinations (including the target area, decoy servers, etc), 
without knowing the attacker's classification, and
\item records sources observed in DoS'ed links to detect suspicious recurring sources.
\end{inparaenum} 

\begin{figure*}[ht]
\begin{center}
\includegraphics[width=1.0\linewidth]{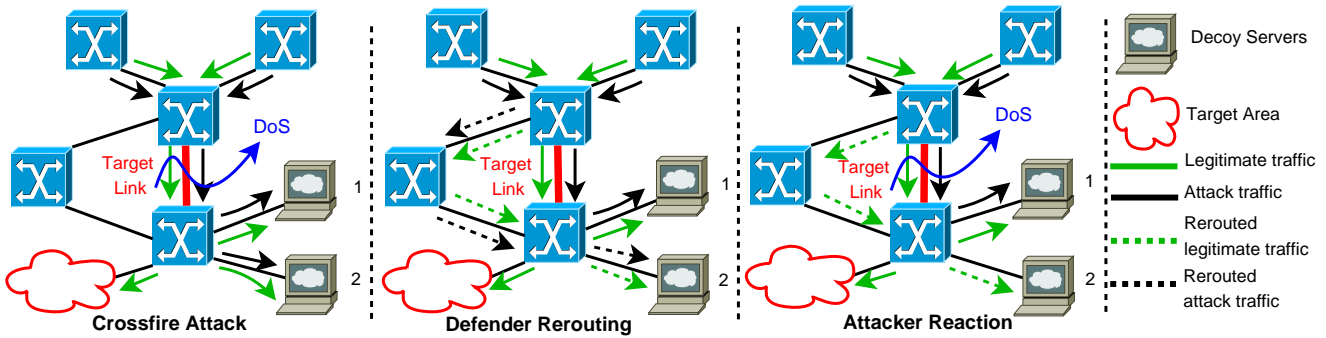}
\vspace{-0.8cm}
\caption{Attacker - Defender interplay: 1) Attacker floods target link, 2) Defender reroutes traffic destined to decoy server 2 to mitigate 
the attack, 3) Attacker updates selected decoy servers sending more traffic to decoy server 1. Traffic sources returning to the target link are deemed suspicious.}
\label{fig:figure}
\end{center}
\vspace{-0.8cm}
\end{figure*}

\textbf{Attacker - Defender Interplay:}
Malicious \textit{Crossfire} bots will change their decoy server selection in case the rerouting has diverted their load
from the targeted link(s), as shown in the example of Figure~\ref{fig:figure}. Therefore, the same bots will ``return to
the crime-scene'' and be  present in another DoS event.
The defender records the sources that appear in DoS'ed links. In addition, he records sources that he re-routes ``away'' of
such links for load balancing. Sources that are pushed away, but return to future DoS'ed links (by selecting a new decoy
server) are particularly suspicious. Thus after each attacker-defender interaction, malicious sources become more identifiable.
Sources  that surpass a threshold of suspiciousness can be rate-limited. In contrast, benign flash-crowd load does not
re-adjust to routing changes, but it uses the same popular destination(s) as before. 

A key challenge for the defender is that he has no information about how the network nodes 
are mapped to the target area and the decoy servers. 
Therefore, he can select different re-routing strategies to optimise the detection phase: 
e.g., using random destinations or preferring servers that occupy homogeneous bandwidth 
on the congested links. 
The former strategy is simpler but might need more rounds to lead to detection. 
The latter strategy is a crude way of ``approximating'' the decoy server groups, since the attacker
generally distributes traffic to those evenly for load-balancing.

\textbf{Parameter Space:} In our model, the attacker and the defender are continuously
engaged into two co-depe-ndent loops consisting of the following steps: 1) centralised flow monitoring~\cite{monitoring};
2) consolidation of measurement data and decision for next action; 3) network control and 
bot orchestration.
The interactions between the two players determine if and when one of them has an
advantage. Numerous parameters can influence the outcome, namely:
\begin{inparaenum}[\itshape i\upshape)]
\item network topology and link capacities,
\item attack flow rate per bot,
\item bot and flow assignment strategy for attacker,
\item link-map and link-bandwidth measurement intervals for both the attacker and the defender,
\item rerouting strategy for the defender,
\item DoS'ed link detection approach and thresholds, and
\item suspiciousness threshold for rate-limiting decisions.
\end{inparaenum}

\textbf{Proof-of-Concept Implementation:} To better understand the parameter space
and the parameterization trade-offs,
we have implemented a full prototype of the attacker and the defender on the Mininet emulation
environment~\cite{mininet}. Emulated bots periodically use traceroute towards the target area and the decoy servers
to build and maintain the link-map. They flood the target links using low-bandwidth HTTP GET requests to the
decoy servers. On the defender's side, we use the POX OpenFlow controller~\cite{pox} for prototyping our online
traffic engineering approach. We also employ OpenFlow~\cite{openflow_1_0} capabilities for network monitoring,
in order to have a unified management solution. More implementation details of our emulation setup can be found
in~\cite{gkounis}.

\textbf{Preliminary Results and Insights:} We evaluate the feasibility of our approach
using a custom topology 
as described in~\cite{gkounis}.
Early results show that our SDN-based solution reacts in comparable time scales
(some seconds) as the attack setup stage. 
Since the attacker needs approx. equal time to identify and react to defender's re-routing
and counter-measures, the attack is effective for 
$\sim$50\% of the total time. 
Moreover, a key insight is that links that are DoS'ed in parallel 
need to be handled in batches to avoid routing oscillations during mitigation; this mandates a tunable delay in the defender's action loop.
We further note that the OpenFlow-based \emph{control} functions run in sub-second times; \emph{monitoring} is 
the dominant time component.

\section{Summary and Open Questions}


In this work, we proposed using dynamic traffic engineering in a novel way to detect and counter the particularly stealthy 
\textit{Crossfire} attack. In contrast to the CoDef~\cite{codef} approach which assumes a collaborative environment,
we assume a hostile setup.
The approach raises a number of research questions:
\begin{inparaenum}[\itshape i\upshape)]
\item What is the trade-off between detection accuracy, topology and traffic characteristics, and re-routing strategy? How can we re-route traffic to accelerate detection
and minimise false positives?
\item What is the trade-off between rerouting costs (including possible instabilities) and DoS costs?
\item What are the temporal aspects of the attacker and defender control loops and their dependencies on different parameters?
\item How to extend the mitigation methodology to the inter-domain level involving support from upstream providers, who run the
algorithm ``as-a-Service''? 
\item How to scale up source recording using bloom filters and/or prefix aggregation?
\end{inparaenum}
Other future research topics include understanding the interplay between fast rerouting and TCP congestion control~\cite{gao2007interactions}, 
as well as the interaction with classic routing policies~\cite{caesar2005bgp}.
Lastly, we are interested in the game-theoretic modelling of the interaction between the players.





%
\vfill\eject
\balance
\bibliographystyle{acm}
{
\bibliography{sigproc}  
}

\end{document}